\documentclass[a4paper, accepted=2025-08-27]{quantumarticle}
\pdfoutput=1
\usepackage{float, verbatim, amsmath, amssymb, ulem, color, hyperref}
\usepackage[numbers,sort&compress]{natbib}

\newcommand{\gate}[1]{\texttt{#1}}

\title{Describing Trotterized Time Evolutions on Noisy Quantum Computers via Static Effective Lindbladians}
\author{Keith R.\ Fratus}
\author{Kirsten Bark}
\author{Nicolas Vogt}
\author{Juha Leppäkangas}
\author{Sebastian Zanker}
\author{Michael Marthaler}
\author{Jan-Michael Reiner}
\affiliation{HQS Quantum Simulations GmbH, Rintheimer Straße 23, 76131 Karlsruhe, Germany}

\begin{document}

\maketitle

\begin{abstract}

We consider the extent to which a Trotterized time evolution implemented on a quantum computer is altered by the presence of decoherence. Given a specific set of assumptions regarding the manner in which noise processes acting on such a device can be modeled at the circuit level, we show how the effects of noise can be reinterpreted as a shift to the dynamics of the original system being simulated.
In particular, we find that this shift can be described through the use of static Lindblad noise terms, which act in addition to the original unitary dynamics. The form of these noise terms depends not only on the underlying noise processes occurring on the device, but also on the original unitary dynamics, as well as the manner in which these dynamics are simulated on the device, i.e., the choice of quantum algorithm.
We call this effectively simulated open quantum system the noisy algorithm model.
Our results are confirmed through numerical analysis.

\end{abstract}

\section{Introduction}
\label{sec:introduction}

Simulating the time evolution of quantum systems is widely discussed as one of the prime applications of quantum computers due to the exponential speedup these devices promise over conventional computers~\cite{feynman_simulating_1982, lloyd_universal_1996, georgescu_quantum_2014}.
Current error rates on present-day universal devices, however, prohibit solving more than small-scale example systems~\cite{martinez_real-time_2016, kandala_hardware-efficient_2017, kandala_error_2019, arute_observation_2020}, while quantum error correction remains out of reach for the foreseeable future~\cite{fowler_surface_2012, devitt_quantum_2013, lekitsch_blueprint_2017}.
As a result, research regarding early utilization of quantum computers often focuses on algorithms with low circuit depth~\cite{bharti_noisy_2022}, and on mitigating errors rather than trying to remove them completely~\cite{RevModPhys.95.045005, russo_testing_2022}.
In this endeavor of enabling useful, near-term quantum computing, it is crucial to understand the effects that noise can have on the results of a simulation performed on such a device. While we have investigated this question already in earlier work for specific noise types and quantum systems~\cite{reiner_effects_2018}, we present here a more extensive approach to this problem.

We focus in this work on the time evolution of quantum spin systems -- systems described by a Hamiltonian in which a number of spin degrees of freedom experience few-body interactions among each other. A wide variety of physical systems are well-approximated by such a description, but solving quantum spin systems is in general hard, either analytically or using conventional computers~\cite{seetharam_digital_2021, doronin_simulation_2021}. Since there exists a direct mapping between spin degrees of freedom and qubits on a quantum device, such a time-evolution can be implemented in a natural fashion on such a device using the Suzuki-Trotter decomposition and the natively available gate set~\cite{lloyd_universal_1996, nielsen_quantum_2010}. However, the presence of noise will result in gate operations which are not faithful representations of their intended unitary operations, and thus in turn will alter the true time evolution of the quantum register.

Our aim in the present work is to understand how the effects of noise on such a time evolution can be interpreted as a shift to the dynamics driving this time evolution. In a separate work \cite{nap}, we have argued for the validity of a particular model for how the effects of noise in a quantum device manifest at the circuit level. Given such a model, we demonstrate that the shift to the dynamics can be well-approximated through the use of static Lindblad noise terms, which act in addition to the existing unitary dynamics. The nature of these Lindblad terms depends on the noise present on the device, but also on the particular choice of Hamiltonian dynamics, as well as the manner in which these Hamiltonian dynamics are implemented on the device as a sequence of gate operations, i.e., the quantum algorithm. For this reason, we call the resulting effective Lindbladian the \textit{noisy algorithm model}.

The outline of this paper is as follows: In Section~\ref{sec:time-evolution-of-spin-systems} we discuss the types of spin systems and quantum algorithms considered in our analysis, and in Section~\ref{sec:noise-assumptions} we outline our assumptions regarding the nature of the noise on the devices we consider. The main results of our analysis, presented in Section~\ref{sec:noisy-algorithm-model}, constitute a method for deriving the noisy algorithm model for a given quantum circuit, along with some of its general properties, while a numerical analysis of its accuracy is given in Section~\ref{sec:numerical-analysis}. We conclude in Section~\ref{sec:conclusion}.
In Appendix~\ref{sec:software-implementation} we describe our software implementation of the presented method, while in Appendix~\ref{sec:appError} we give an extensive error analysis of our methods.

\section{Digital Simulation of Spin Systems}
\label{sec:time-evolution-of-spin-systems}

We will restrict ourselves to quantum circuits which involve the digital simulation of the time evolution of a quantum system comprised of a set of spin-$\frac{1}{2}$ degrees of freedom (or simply ``spins''), with the Hamiltonian
\begin{equation}
H = \sum_{X} h_{X}.
\end{equation}
Each subset $X$ is limited to a non-extensive number of spins (in practice two spins at most), and $h_{X}$ describes the interactions among these spins. For a system with $k$ spins, the Hilbert space $\mathcal{H}$ has dimension $\mathcal{D}=2^k$. For simplicity, we will assume that the Hamiltonian is time-independent, although the generalization of our results to the time-dependent case is straightforward. A common example of such a Hamiltonian would be the Transverse-Field Ising Model with nearest-neighbor interactions,
\begin{equation}\label{eq:ising-model}
H = J\sum_{\langle i j \rangle} \sigma^{z}_{i} \sigma^{z}_{j} + g \sum_{i} \sigma^{x}_{i},
\end{equation}
the physics of which depends strongly on the geometry of the underlying lattice. Since the degrees of freedom in the Hamiltonian are spin-$\frac{1}{2}$ observables, it is possible to associate each degree of freedom with a qubit on a quantum device, without the need for additional transformations (for example, the Jordan-Wigner transformation in the case of fermionic degrees of freedom).

To perform such a digital simulation, the unitary time evolution is approximated using the usual Trotter expansion~\cite{trotter_product_1959, suzuki_generalized_1976, hatano_finding_2005},
\begin{equation}
\begin{split}
U \left ( t \right ) & = \exp \left ( - i H t \right ) = \prod_{n=1}^{N} \exp \left ( - i H \tau \right )  \\
& \approx \prod_{n=1}^{N} \prod_{X}  \exp \left ( - i h_{X} \tau \right ) \equiv \prod_{n=1}^{N} \prod_{X}  U_{X} \left ( \tau \right )
\end{split}
\end{equation}
where $\tau = t / N$. Since the individual terms $h_{X}$ in the Hamiltonian each involve only a small number of sites, the unitary operators appearing in the product on the right can be efficiently simulated using gates natively available on a quantum device~\cite{lloyd_universal_1996, nielsen_quantum_2010}. Such a Trotter expansion of course involves some degree of approximation. The effective Hamiltonian which is actually being evolved in such a simulation can be found by using the \textit{Baker-Campbell-Hausdorff} (BCH) formula to recombine the product into a single exponential, which to first order in the Trotter step size yields
\begin{equation}
H \to H_{\text{eff}} = H + \delta H = \sum_{X} h_{X} - \frac{i}{2} \tau \sum_{X < Y} \left [ h_{X}, h_{Y} \right ],
\end{equation}
where $X < Y$ when the term $h_{X}$ appears to the left of $h_{Y}$ in the Trotter product. Our analysis will generally neglect this first-order correction to the Hamiltonian, though we will study the accuracy of this choice later.

\section{Noise in Quantum Circuits}
\label{sec:noise-assumptions}

Throughout the implementation of a Trotter step, decoherence will lead to the accumulation of errors in our simulation. Unlike the errors introduced through Trotterization, however, these errors lead to non-unitary evolution of the qubit register. In order to analyze the effects of this noise, we must assume a model for how noise manifests itself at the circuit level.

\subsection{The Lindblad Equation}

To begin accounting for the effects of noise in our simulation in concrete terms, we must adopt the notation of the density matrix, rather than the pure wave function, which cannot describe the evolution of mixed quantum states. In the strictly unitary case in which the density matrix remains pure, this replacement can be made according to
\begin{equation}
| \psi \left ( t \right ) \rangle ~\to~ \rho \left ( t \right ) = | \psi \left ( t \right ) \rangle \langle \psi \left ( t \right ) |,
\end{equation}
so that the dynamics of the density matrix are given according to
\begin{equation}
\begin{split}
\rho \left ( t \right ) & = \exp \left ( - i H t \right ) | \psi \left ( 0 \right ) \rangle  \langle \psi \left ( 0 \right ) |  \exp \left ( + i H t \right ) \\ & = 
 \exp \left ( - i H t \right ) \rho \left ( 0 \right ) \exp \left ( + i H t \right )
\end{split}
\end{equation}
Using the identity,
\begin{equation}
e^{\text{ad}_{X}} Y =  e^{X}Ye^{-X}~;~\text{ad}_{X} \equiv \left [ X, \cdot \right ], 
\end{equation}
the expression for the time-evolution of the density matrix can be further rewritten as,
\begin{equation}
\rho \left ( t \right ) = e^{\mathcal{L}_{H} t} \rho_{0},
\end{equation}
where the Liouvillian super-operator for the time-evolution of the density matrix is a linear operator acting on the space of density matrices, given as
\begin{equation}
\mathcal{L}_{H} \equiv -i ~ \text{ad}_{H} = -i \left [ H, \cdot \right ]
\end{equation}

Moving beyond the unitary case, the density matrix will, in general, no longer remain pure,
\begin{equation}
\text{Tr}\left[\rho^{2} \left ( t \right )\right] < 1.
\end{equation}
However, in order to respect the basic statistical interpretation of quantum mechanics, the density matrix must remain a positive semi-definite matrix with unit trace. In other words, the time-evolution of the density matrix must be a \textit{completely positive, trace-preserving} (CPTP) map. If, in addition to this basic requirement, we also require the dynamics to be linear, Markovian, and time-homogeneous, then the most general map on the space of density matrices which meets our requirements is generated by the so-called  \textit{Lindblad equation} \cite{lindblad, 10.1063/1.522979},
\begin{equation}
\begin{split}
\mathcal{L} \left [ \rho  \right ] & = \mathcal{L}_{H}\left [ \rho  \right ] + \mathcal{L}_{D}\left [ \rho  \right ] = \\
 -i \left [ H, \rho \right ] & + \sum_{n, m} \Gamma_{nm} \left ( A_{n} \rho A_{m}^{\dagger} - \frac{1}{2} \left \{ A_{m}^{\dagger} A_{n}, \rho \right \}  \right )
 \end{split}
\end{equation}
The first term in this equation is recognizable as the original unitary evolution, while the second piece, often referred to as the Lindblad term, accounts for decoherence through noise. The operators $ \left \{ A_{n} \right \} $ represent a basis for the space of all traceless operators on $\mathcal{H}$, while the time-independent \textit{rate matrix} $\Gamma$ must be Hermitian and positive semi-definite in order to preserve the statistical properties of the density matrix. If we relax the assumption that the rate matrix is time-independent, such that the dynamics are no longer time-homogeneous, then the necessary conditions for preserving the statistical interpretation of the density matrix are still not fully understood \cite{RevModPhys.88.021002}. We will not concern ourselves with this case here.

The set of operators $\left \{ A_{n} \right \}$ is not unique - any basis of traceless operators is valid. For our purposes, however, we will restrict ourselves to a basis which is orthonormal with respect to the usual Frobenius inner product on matrices,
\begin{equation}
\frac{1}{\mathcal{D}}\text{Tr} \left [ A^{\dagger}_{m}A_{n}\right ]  ~=~ \delta_{mn}.
\end{equation}
We find such a basis to be ideal for stating our results cleanly, though other choices of normalization are possible (it is however important to remain consistent in this choice, since the value of the rate matrix will depend on it). Common choices for such a basis $ \left \{ A_{n} \right \} $ include the (properly normalized) generalized Gell-Mann matrices \cite{Bertlmann_2008}, or, since we are working with collections of qubits, the set of all possible products of Pauli operators on $k$ qubits,
\begin{equation}
P_{\alpha} \equiv \bigotimes_{i=1}^{k} \sigma_i^{\alpha_i} ~;~ \alpha_{i} \in \left \{  0, x, y, z \right \}
\end{equation}
For a given Lindblad term, changing the basis of traceless operators will result in a corresponding change in the form of the rate matrix, according to a transformation law which can be found in \cite{nap}.

One particularly relevant choice of (orthonormal) basis is given by
\begin{equation}
L_{i} = \sum v_{i}^{(n)} A_{n}.
\end{equation}
where the vectors $\left \{ \vec{v}_{i} \right \}$ are the (normalized) eigenvectors of the rate matrix (as expressed in the basis $ \left \{ A_{n} \right \} $). Because the rate matrix is Hermitian, these eigenvectors will always constitute a complete basis with real eigenvalues $\{\gamma_{i}\}$, and since the rate matrix is positive semi-definite, these eigenvalues will be non-negative. This choice of basis leads to the more common diagonal form of the Lindblad term,
\begin{equation}
\mathcal{L}_{D}\left [ \rho  \right ] = \sum_{i} \gamma_{i} \left [ L_{i}  \rho L_{i}^{\dagger} - \frac{1}{2} \left \{ L_{i}^{\dagger} L_{i}, \rho \right \} \right ],
\end{equation}
The eigenvalues $\{\gamma_{i}\}$ correspond to the physical decay rates of the system under the effects of decoherence.

Many common examples of Lindblad noise which are often encountered in the literature involve uncorrelated decoherence acting on individual qubits. These include damping noise,
\begin{equation}
\mathcal{L}_{\text{damp}}\left [ \rho \right ] = \sum_{j} \gamma_{j}^{\text{damp}} \left [ \sigma_{j}^{+} \rho \sigma_{j}^{-} - \frac{1}{2} \left \{ \sigma_{j}^{-} \sigma_{j}^{+}, \rho \right \} \right ],
\end{equation}
dephasing noise,
\begin{equation}
\mathcal{L}_{\text{deph}}\left [ \rho \right ] = \frac{1}{2}\sum_{j} \gamma_{j}^{\text{deph}} \left [ \sigma_{j}^{z} \rho \sigma_{j}^{z} - \rho \right ],
\end{equation}
and depolarizing noise,
\begin{equation}
\mathcal{L}_{\text{depo}}\left [ \rho \right ] = \frac{1}{4}\sum_{j} \gamma_{j}^{\text{depo}} \sum_{\alpha \in \left \{ x, y, z \right \}} \left [ \sigma_{j}^{\alpha} \rho \sigma_{j}^{\alpha} - \rho \right ].
\end{equation}
The rates $\left \{ \gamma_{j}^{\text{damp}} \right \}$, $\left \{ \gamma_{j}^{\text{deph}} \right \}$, and $\left \{ \gamma_{j}^{\text{depo}} \right \}$ indicate the characteristic noise strength at each site in the system, and aside from some conventional numerical factors, coincide with the eigenvalues of the rate matrix (which is of course simply diagonal in these cases). For example, for an observable $\mathcal{O}_{X}$ which is a product of Pauli operators living on the sites in the set $X$, evolving under purely uncorrelated depolarizing dynamics, we have
\begin{equation}
\frac{d}{dt} \langle \mathcal{O}_{X} \rangle = - \left ( \sum_{j \in X} \gamma_{j}^{\text{depo}} \right ) \langle \mathcal{O}_{X} \rangle.
\end{equation}
One should note, however, that the conventional choice of basis for damping noise satisfies
\begin{equation}
\frac{1}{\mathcal{D}}\text{Tr} \left [ \sigma^{+}_{i} \sigma^{-}_{j} \right ] = \frac{1}{2}\delta_{ij},
\end{equation}
and so care must be taken when properly defining the rate matrix in this case. More discussion of the Lindblad equation can be found in \cite{lidar, breuer2002theory}.

\subsection{Modeling Noise at the Circuit Level}

With this understanding of the Lindblad equation in mind, we now outline our model for how the effects of noise manifest themselves at the circuit level. Our assumptions throughout this work regarding noise will be based upon a separate work~\cite{nap} which concludes that the effects of noise can be accounted for at the level of individual gates. To be more precise, the conclusion of this work is that a circuit can be modeled as a sequence of noise-free quantum gates, $\left\{G\right\}$, with each gate followed by an individual, discrete decoherence event $\mathcal{N}_{G}$. These decoherence events can be adequately described as Lindblad noise accumulating during a small but finite time duration, namely the gate application time,
\begin{equation}
\mathcal{N}_{G} \to \exp \left ( t_{G} \mathcal{L}^{G}_{N} \right ),
\end{equation}
where $\mathcal{L}^{G}_{N}$ represents a pure Lindblad noise term (i.e., there is no Hamiltonian component). Further insight into how $\mathcal{L}^{G}_{N}$ can be calculated in specific situations can be found in~\cite{nap}.

We emphasize that the form of $\mathcal{L}^{G}_{N}$ may differ significantly from the form of the underlying environmental processes acting directly on the quantum register, and the relationship between the two will depend upon the precise manner in which the quantum gate is implemented as a sequence of operations on the register. For example, $\mathcal{L}^{G}_{N}$ may involve multi-qubit correlated noise, even if the underlying noise processes are uncorrelated. However, under the set of assumptions outlined in~\cite{nap} regarding how these operations act, $\mathcal{L}^{G}_{N}$ can always be written in Lindblad form, at least to first order in the noise strength. While these assumptions will not be valid for all sources of error on all forms of quantum hardware~\cite{PhysRevLett.116.020501, PhysRevLett.121.090502, PhysRevLett.123.190502, PhysRevLett.117.060504, RevModPhys.87.1419, Saffman_2016, cardani_reducing_2021}, they are nevertheless well established in the community and proven to be reasonably accurate in practice~\cite{lidar}.

While the form of the discrete noise following a gate will generally depend on the particular choice of gate, and the particular choice of hardware, we will assume that the form of this noise is known (perhaps as a result of tomography performed on the relevant device~\cite{boulant_robust_2003,howard_quantum_2006,samach_lindblad_2022}), and that it remains constant throughout at least a single run. We will also assume that the noise event which occurs after a gate affects only those qubits which are involved in the gate - in other words, the noise following a gate operation on a set of qubits does not lead to any entanglement with any of the other qubits on the device. Note that this framework includes the noise accumulated on qubits as they idle, when accounting for the action of the ``trivial'' gate (in other words, the action of doing nothing to a qubit).

Having chosen a model for how the effects of noise manifest themselves at the level of a quantum circuit, we now proceed to analyze how this noise alters the effective model simulated by the circuit.

\section{The Noisy Algorithm Model}
\label{sec:noisy-algorithm-model}

Over the course of a single Trotter step, the state of the quantum register will have evolved from some initial $\rho$, to some final $\rho'$. In the ideal case in which there is no Trotter error or decoherence, this evolution would correspond to unitary dynamics under the originally desired Hamiltonian,
\begin{equation}
\rho ' = e^{\mathcal{L}_{H} \tau } \rho
\end{equation}
However, the discrete noise terms occurring in the quantum circuit (as well as the errors resulting from Trotterization) will result in a time-evolution which deviates from this ideal case. Our aim in this work is to describe these effects through an effective time-evolution operator
\begin{equation}
\begin{split}
\rho ' & = e^{\mathcal{L_{\text{eff}}} \tau } \rho  ~;~ 
\mathcal{L}_{\text{eff}}  \left [ \rho \right ] \equiv \\
 - i \left [ H^{\text{eff}}, \rho \right ] & + \sum_{n, m} \Gamma^{\text{eff}}_{nm} \left ( A_{n} \rho A_{m}^{\dagger} - \frac{1}{2} \left \{ A_{m}^{\dagger} A_{n}, \rho \right \}  \right ),
\end{split}
\end{equation}
and thereby interpret the quantum circuit as now performing a simulation of this effective model, the \textit{noisy algorithm model}, rather than the original coherent model. We now proceed to characterize $\mathcal{L}_\mathrm{eff}$, the principal component of the noisy algorithm model.

\subsection{Circuits with Native Gates}

To begin our analysis, we will assume that all of the exponential products in the Trotter expansion correspond to natively available gates on the given hardware. In other words, we will assume that the unitary operators
\begin{equation}
U_{X} \left ( \tau \right ) \equiv \exp \left ( - i h_{X} \tau \right )
\end{equation}
are natively available on the hardware as quantum logic gates (we will relax this assumption shortly). Written as a super-operator acting on the space of operators,
\begin{equation}
\mathcal{U}_{X} \left ( \tau \right ) \equiv e^{\mathcal{L}_{X} \tau} ~;~ \mathcal{L}_{X} \equiv - i~ \text{ad}_{h_{X}} = -i \left [ h_{X}, \cdot \right ]
\end{equation}

The Hamiltonian term $h_{X}$ will generally consist of a term proportional to a product of Pauli operators, for example,
\begin{equation}
h_{X} = J_{X} \sigma_{i}^{z} \sigma_{j}^{z},
\end{equation}
where sites $i$ and $j$ live on the domain $X$. For the Trotter decomposition to be well-behaved, we must generally assume that the quantity
\begin{equation}
\phi_{X} = 2 J_{X} \tau
\end{equation}
is sufficiently small. Thus, $U_{X}$ corresponds to the exponentiation of an argument which is parametrically small in the quantity $\phi$. We will refer to such an operation as a \textit{small angle gate}~(SAG).

All of these gate operations will of course have non-unitary dissipation terms interspersed among them. These terms also correspond to the exponentiation of an argument which is parametrically small in some quantity. In this case, however, the small quantity is the product of the gate time with the characteristic noise strength,
\begin{equation}
\mu_{X} = \gamma_{X} t_{X}.
\end{equation}
For example, for independent dephasing noise on each of the two qubits on $X$ following the application of the above gate, the argument of the exponential term will be
\begin{equation}
\begin{split}
& t_{X}\mathcal{L}^{X}_{N}  \left [ \rho \right ] \\ = ~ & \frac{1}{2} \gamma_{i}^{\text{deph}} t_{X}  \left ( \sigma_{i}^{z} \rho \sigma_{i}^{z} -  \rho \right ) + \frac{1}{2} \gamma_{j}^{\text{deph}}  t_{X} \left ( \sigma_{j}^{z} \rho \sigma_{j}^{z} -  \rho \right ) \\ \equiv ~ & \frac{1}{2} \mu_{i}^{\text{deph}}   \left ( \sigma_{i}^{z} \rho \sigma_{i}^{z} -  \rho \right ) + \frac{1}{2} \mu_{j}^{\text{deph}}   \left ( \sigma_{j}^{z} \rho \sigma_{j}^{z} -  \rho \right )
\end{split}
\end{equation}
In general, we will assume that there are some characteristic $\phi$ and $\mu$ which describe the coherent and incoherent terms throughout the circuit, respectively, and that they satisfy
\begin{equation}
\mu \ll \phi
\end{equation}
Without this assumption, the effects of noise on the device would be too great to perform any sort of useful computation.

Despite not being unitary, the noise terms are still generated through exponentiation of some linear (super) operator, and hence it is still possible to combine a noise term and a coherent gate term into a single exponential,
\begin{equation}
e^{t_{G} \mathcal{L}^{G}_{N}}e^{\tau \mathcal{L}_{X}} \to e^{\tau \mathcal{L}},
\end{equation}
where $\tau \mathcal{L}$ is given according to the BCH formula. Since all of the noise and gate terms appearing in the circuit correspond to the exponentiation of ``small'' quantities, to lowest order in the BCH expansion we can combine all of these terms into a single exponential, by merely summing up all of their arguments,
\begin{equation}
\exp \left ( \tau \mathcal{L}_{\text{eff}}  \right ) \approx \exp \left (\tau \sum_{X} \mathcal{L}_{X} + \sum_{g} t_{g} \mathcal{L}^{g}_{N} \right ),
\end{equation}
where the first summation is over all SAGs in the circuit, and the second summation is over all noise terms in the circuit (in this case where we only consider SAGs, these summations are in direct correspondence with each other). Dividing through by the Trotter step size $\tau$, we find
\begin{equation}
\mathcal{L}_{\text{eff}} \approx \sum_{X} \mathcal{L}_{X} + \sum_{g} \left ( t_{g} / \tau \right )\mathcal{L}^{g}_{N}.
\end{equation}

The first term in this expression naturally corresponds to the originally desired time evolution. Since we have only recombined the exponential terms to lowest order, we do not see any corrections to the Hamiltonian which come from Trotter error. While we will generally neglect such corrections due to Trotterization in the remainder of our analysis, as our primary interest here concerns corrections due to noise, we will nevertheless return to these corrections in Section~\ref{sec:scale}, where we consider the scaling of terms which have been neglected in our analysis. The second term above represents the cumulative effects of all of the discrete noise terms occurring in the circuit. Thus, we find that we can interpret the effects of decoherence on the device as introducing an additional dissipative term to our simulated model,
\begin{equation}
\mathcal{L}_{H} =  -i \left [ H, \cdot \right ] ~ \to ~ \mathcal{L}_{\text{eff}} \approx  -i \left [ H, \cdot \right ] + \mathcal{L}_{N}
\end{equation}
where
\begin{equation}\label{eq:noise-rescaling}
\mathcal{L}_{N} ~ = ~ \sum_{g} \left ( t_{g} / \tau \right ) \mathcal{L}^{g}_{N}.
\end{equation}

The most prominent feature of this result is that the contribution to the effective noise from each discrete noise term depends on the ratio of gate time to simulated Trotter step size. This is of course reasonable, since the longer it takes to implement a single Trotter step, the more decoherence will accumulate on the qubit register. When interpreting the evolution of the qubit register in the context of a simulated model with a fixed simulation time, a larger total decoherence will correspond to an effective model with a larger decoherence rate.

\subsection{Complications from ``Large'' Gates}
\label{sec:large-gates}

So far, we have assumed that all gate operations occurring in our circuit correspond directly to a term appearing in the Trotter expansion, $U_{X} \left (\tau \right )$, parameterized by some small quantity. However, in reality, many of these terms $U_{X}$ are not natively available as quantum gates on the hardware level, and must be decomposed into a set of available gates.

\begin{figure}
   \centering
   \includegraphics[width=\columnwidth]{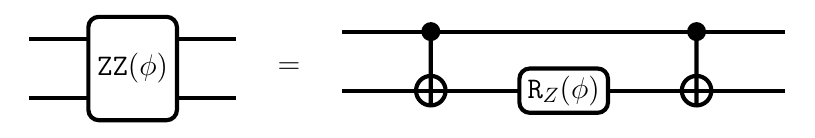}
   \caption{Demonstrating the idea of a block decomposition: A \gate{ZZ} Ising gate can be decomposed into \gate{CNOT} gates and a single qubit rotation around the $Z$ axis.}
   \label{fig:block}
\end{figure}

\begin{figure}
   \centering
   \includegraphics[width=\columnwidth]{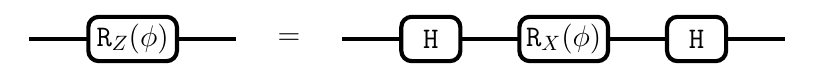}
   \caption{Another example of a gate decomposition: Using Hadamard gates to effectively change the axis of a rotation gate.}
   \label{fig:had}
\end{figure}

An example would be an Ising interaction term $J \sigma^{z}_{i} \sigma^{z}_{j}$ from the Hamiltonian~\eqref{eq:ising-model}, where the exponential in the Trotter expansion can be implemented through a \gate{ZZ} Ising gate via
\begin{equation}
e^{-\mathrm{i} J \sigma^{z}_{i} \sigma^{z}_{j} \tau} = e^{-\mathrm{i} \frac{\phi}{2} \sigma^{z}_{i} \sigma^{z}_{j}} = \gate{ZZ}(\phi),
\end{equation}
with the angle $\phi = 2J\tau$. The \gate{ZZ} gate can be decomposed using \gate{CNOT}s and a rotation gate around the $Z$ axis:
\begin{equation}
\gate{ZZ}(\phi) = \gate{CNOT}_{ij} \cdot \gate R_{Z_j} (\phi) \cdot \gate{CNOT}_{ij},
\end{equation}
where $i$ is the control qubit, and $j$ is the target qubit of $\gate{CNOT}_{ij}$ and the rotation is acting on the target qubit. Such a decomposition is illustrated in Figure~\ref{fig:block}. 
Another example is the use of Hadamard gates to modify a rotational axis,
\begin{equation}
\gate{R}_Z(\phi) = \gate H \cdot \gate R_X (\phi) \cdot \gate H,
\end{equation}
illustrated in Figure~\ref{fig:had}. The key feature of these decompositions is that they involve gate operations (here, for example, \gate{CNOT} or Hadamard gates) which \textit{cannot} be written as the exponentiation of some small term. We will therefore refer to such a sequence of gates as a \textit{large gate decomposition block}~(LGDB).

Since LGDBs contain terms which do not correspond to the exponentiation of some small parameter, simply summing together the exponential terms to first order in the BCH expansion is no longer a valid approximation. If we wish to analyze the effects of noise in circuits containing LGDBs using the techniques of the previous section, we must map our circuit to an equivalent one in which only small-angle Hamiltonian terms or noise terms exist. By equivalent circuit, we mean any circuit which results (within the Trotter approximation) in precisely the same evolution of the qubit register as the circuit which is actually being implemented on the quantum device, whether or not such a circuit could actually be implemented in practice. In the case at hand, such an equivalent circuit can be found by identifying any noise terms occurring within a LGDB, and then shifting them to the outside of the block, while accounting for the effects of commuting these noise terms past any coherent gates within the block (we emphasize that since noise is something which an experimenter generally has very little control over, this process of shifting noise terms is simply a mathematical trick for the purposes of analyzing the circuit, and does not correspond to any actual change to the original circuit). Since the original LGDB, by definition, can be recombined back into a small angle term $U_{X} \left (\tau \right )$, we can then proceed with the original analysis using this equivalent circuit. The goal, then, is to determine how the noise terms appearing within one of these LGDBs are modified by the process of commuting them to the outside of the block.

\begin{figure}
   \centering
   \includegraphics[width=\columnwidth]{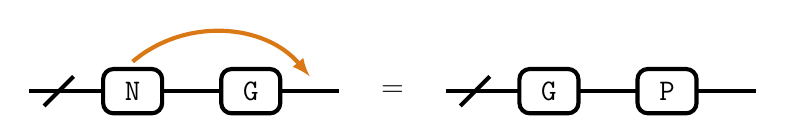}
   \caption{Demonstrating the idea of commuting noise past a gate \gate{G}: The original noise term \gate{N} that acted prior to \gate{G} is replaced by a noise term \gate{P} acting after \gate{G}, where both gate sequences are equivalent. How to find \gate P, given \gate G and \gate N, is explained in the main text. Note that these operations may act on one or multiple qubits, depicted through the dash in the horizontal qubit lines, indicating a register.}
   \label{fig:noiseHop}
\end{figure}

In concrete terms, we would like to solve the equation
\begin{equation}
e^{\mathcal{L}_{G}}e^{\mathcal{L}_{N}} = e^{\mathcal{L}_{P}}e^{\mathcal{L}_{G}},
\end{equation}
where $\mathcal{L}_{G}$ describes the action of a noise-free gate, $\mathcal{L}_{N}$ describes the original noise term, and $\mathcal{L}_{P}$ describes the modified noise term we wish to find. This corresponds to moving a noise term in the circuit diagram from the left to the right of a gate, and is represented schematically in Figure~\ref{fig:noiseHop}. In \cite{nap} we derive a transformation which allows us to find the underlying rate matrix $\Gamma^{P}$ of $\mathcal{L}_{P}$ in terms of the rate matrix $\Gamma^{N}$ of $\mathcal{L}_{N}$. This transformation is given according to
\begin{equation}
\Gamma^{P} = M\Gamma^{N}M^\dagger,
\end{equation}
with the matrix $M$ defined as
\begin{equation}
M_{mn} = \frac{1}{\mathcal{D}}\text{Tr} \left [ A^{\dagger}_{m}U_{G} A_{n} U_{G}^{\dagger} \right ].
\end{equation}
Here the matrix $U_{G}$ is the unitary matrix corresponding to the gate $G$, and the $\left\{A_{n}\right\}$ are again the complete basis of traceless operators in the definition of the Lindblad equation. So long as the basis $\left\{A_{n}\right\}$ is orthonormal, the matrix $M$ will be unitary. As a result, the decay spectrum of a given noise term will be preserved when it is commuted past a unitary gate. In the case that a noise term must be commuted past several gates, the transformation involves the matrix $U$ which is the product of all of the corresponding unitary gate matrices,
\begin{equation}
U \equiv \prod_{g}U_{g},
\end{equation}
which is of course simply the claim that multiple unitary gates in succession correspond to one single unitary gate.

In addition to finding the modified noise term through a transformation which alters the rate matrix but leaves the basis $\left\{A_{n}\right\}$ unchanged, it is also possible to find the modified noise term through a transformation which leaves the rate matrix unchanged, and instead alters the basis of tracless operators. Such a change of basis takes the form
\begin{equation}\label{eq:noise-basis-transformation}
A_{n} \to B_{n} = U A_{n} U^{\dagger}.
\end{equation}
Since the matrix $U$ is unitary, this new basis is again orthnormal. This alternative picture may be useful in some computational contexts. For example, we use this formula in our software implementation to calculate the effective Lindblad terms, as mentioned in Appendix~\ref{sec:software-implementation}.

With this information, we can now pass all noise terms to the outside of their corresponding LGDBs, and are left with a circuit containing only small angle gate terms and noise terms, allowing us to proceed with the original analysis, yielding
\begin{equation}
\mathcal{L}_{H} =  -i \left [ H, \cdot \right ] ~ \to ~ \mathcal{L}_{\text{eff}} \approx  -i \left [ H, \cdot \right ] + \mathcal{L}_{P}
\end{equation}
where now
\begin{equation}\label{eq:noise-resclaing-P}
\mathcal{L}_{P} ~ = ~ \sum_{g} \left ( t_{g} / \tau \right ) \mathcal{L}^{g}_{P}.
\end{equation}
A noise term $\mathcal{L}^{g}_{P}$ appearing in the summation above may in fact correspond to one of the original $\mathcal{L}^{g}_{N}$ when such a noise term need not be commuted past any large gates.

\subsection{Some Additional Complications}

So far, we have analyzed circuits with either small angle gate operations, or LGDBs which could be reduced to such small angle gate operations. Yet, there is a variety of circuits which cannot be reduced to such a paradigm. While a full treatment of all such cases is beyond the scope of this work, we focus here on two particularly prominent cases - the use of \gate{SWAP} gates to accommodate limited connectivity, and the cancellation of gates during circuit optimization.

\subsubsection{Handling \gate{SWAP} Gates}
\label{sec:SWAP-handling}

\begin{figure}
   \centering
   \includegraphics[width=\columnwidth]{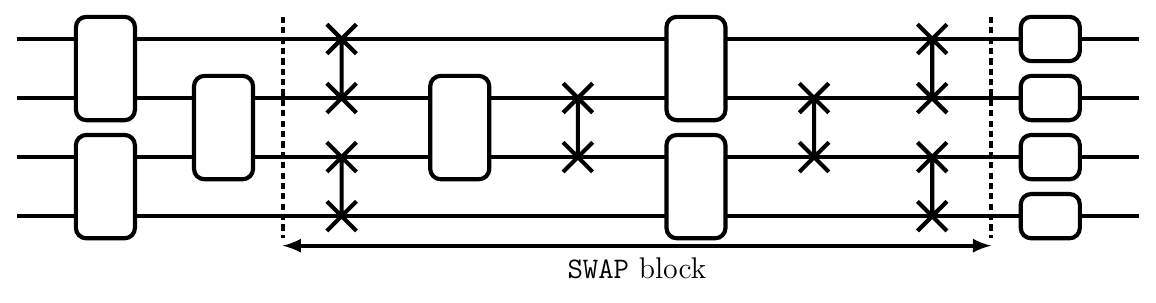}
   \caption{An example of a basic \gate{SWAP} block, where between the swapping operations the qubit indices are effectively scrambled. This is to illustrate that treating \gate{SWAP} blocks similarly to the decomposition blocks in Section~\ref{sec:large-gates} would easily lead to very large blocks, potentially causing a substantial computational overhead commuting all noise terms out of the block.}
   \label{fig:swap}
\end{figure}

An example of a circuit which uses \gate{SWAP} gates to accommodate limited connectivity is shown in Figure~\ref{fig:swap}. Here, we imagine that the four spins possess some pair interactions amongst themselves (the exact nature of this interaction is unimportant). If we assume only linear connectivity among the qubits on our chosen architecture, then the use of \gate{SWAP} gates will be necessary to implement the pair interaction between spins which are not represented by adjacent qubits. This is indicated in the circuit diagram. We also imagine that some other gate operations occur which are designed to implement additional single-qubit Hamiltonian terms. Although not explicitly indicated on the diagram, the pair interaction terms may need to be decomposed into LGDBs. As usual, this case can be handled by commuting the noise terms within the block to the outside, as described previously. However, the \gate{SWAP} gates themselves still pose a problem. If we follow the philosophy of the previous analysis, we must somehow manipulate our circuit so that only small angle gate operations and noise terms exist.

Of course, since the purpose of the \gate{SWAP} gates is to account for pair interactions between non-adjacent spins, we know that, at least in the noise-free case, the circuit is equivalent to one in which the \gate{SWAP}s are removed, and there exist pair interaction terms between the non-adjacent spins. It is clear then that this case can, at least in principle, be handled in a fashion similar to the previous case - one must simply commute out all of the noise terms which occur amid the various \gate{SWAP} gates, so that the operations between the first and last \gate{SWAP} gates can then be combined back into a sequence of operations which only contain small angle operations. In some sense, we can imagine the sequence of gates in between and including these \gate{SWAP} gates as one combined decomposition block. In general, we will refer to such a sequence of gates as a \textit{\gate{SWAP} block}. In practice, however, computing one large transformation matrix $M$ for all of the gate operations occurring within a \gate{SWAP} block may become computationally demanding, especially for larger circuits. Even for the case of four spins, the dimensionality of the rate matrix implies that such transformations can generally be quite computationally demanding.

Fortunately, handling \gate{SWAP} blocks in such a way is not necessary -- it is in fact possible to account for the effects of commuting a noise term outside of a \gate{SWAP} block, without naively computing one giant transformation matrix, by splitting the block into LGDBs and \gate{SWAP}s.

\begin{figure}
   \centering
   \includegraphics[width=\columnwidth]{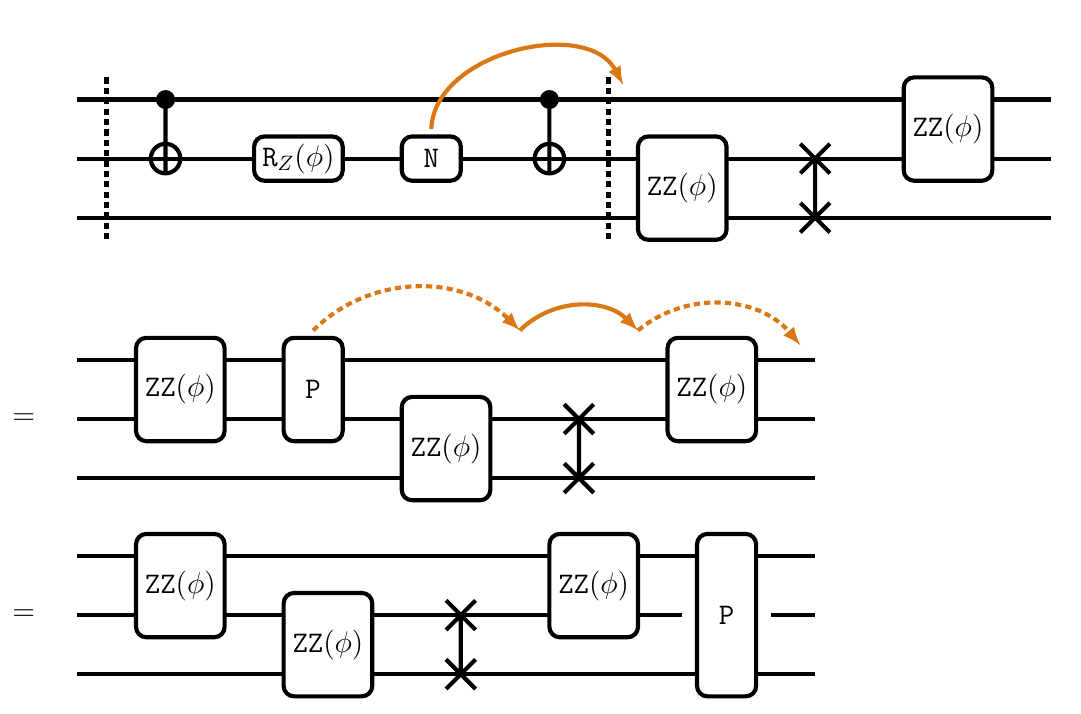}
   \caption{The two-step noise procedure for \gate{SWAP} gates. First, noise terms appearing within a LGDB within the circuit need to be commuted out of the block. Here, this is indicated by the orange arrow in the first line: The noise \gate{N} within the decomposition block between the dashed lines representing a $\gate{ZZ}$ Ising gate (see Figure~\ref{fig:block}) is shifted out of the block. After commuting \gate{N} past the \gate{CNOT} gate we are left with the (in general two-qubit) noise term \gate{P} (similar to Figure~\ref{fig:noiseHop}). Now, the circuit consists only of LGDBs equivalent to small angle gates without noise, separate noise terms, and \gate{SWAP}s. The second step is to commute \gate{P} past all \gate{SWAP}s. When doing this, moving past a LGDB (dashed orange arrows in the second line) does not modify \gate{P}, moving past a \gate{SWAP} gate (solid orange arrow in the second line) will change the qubits \gate{P} is acting on.}
   \label{fig:twoStep}
\end{figure}

To see why, we first note that commuting a noise term past a small angle gate, to lowest (zero) order in the gate angle $\phi$, does not modify the noise (we will address this point in more detail shortly). Second, we note that the effect of commuting a noise term past a \gate{SWAP} gate is to simply exchange the qubits participating in the noise term according to the \gate{SWAP} operation. For this reason, the effects of commuting noise outside of (or past) a \gate{SWAP} block can be accounted for in a two-step process. First, all noise terms occurring within an LGDB are commuted out of the LGDB, resulting in a circuit which now contains only noise, \gate{SWAP} gates, and small angle gate operations. Second, all noise terms are commuted past the \gate{SWAP} gates. In this second step, the only modifications to the noise terms we must account for are the qubit permutations which arise due to the \gate{SWAP} gates, which does not require the computation of a transformation matrix $M$. Such a two-step process is indicated schematically in Figure~\ref{fig:twoStep}.

\begin{figure}
   \centering
   \includegraphics[width=\columnwidth]{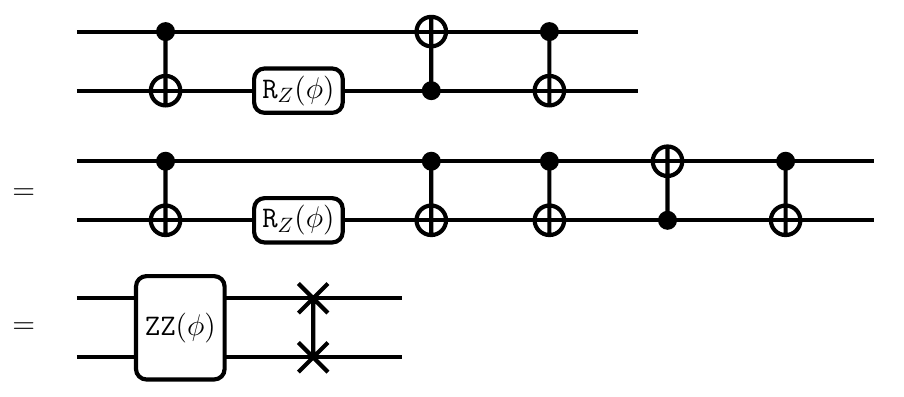}
   \caption{By adding an identity operation in the form of two \gate{CNOT} gates into the gate sequence depicted at the top, one can find that this sequence is in fact equal to a \gate{ZZ} Ising interaction as shown in Figure~\ref{fig:block}, followed by a \gate{SWAP} gate (which can be decomposed into three alternating \gate{CNOT}s).}
   \label{fig:phantom}
\end{figure}

Lastly, we comment on the need to account for swapping operations which occur without the use of explicit \gate{SWAP} gates. Such a case is displayed in Figure~\ref{fig:phantom}. In the noise-free case, such a sequence of gates is equivalent to a pair-Z interaction, followed by a \gate{SWAP} gate. Therefore, in order to treat this case, we must commute any noise terms to the outside of this gate sequence, replace the relevant gates with an effective small angle pair interaction followed by a \gate{SWAP} gate, and then proceed with the previous analysis for handling \gate{SWAP} blocks.

The issue of implicit swapping operations, as well as the fact that on quantum devices usually \gate{SWAP} gates need to be decomposed into native hardware gates (e.g., three alternating \gate{CNOT}s as in Figure~\ref{fig:phantom}), needs to be accounted for when analyzing circuits through software. Our software implementation circumvents these complications by adding an optional property to LGDBs that represent parametrically small gates within a Trotter step: A permutation, that accounts for implicit swapping within the LGDB. Commuting noise terms past a LGDB does not alter the noise type (which is valid within the Trotter approximation); it only alters the qubit indices the noise acts on, according to the given permutation. More details on the software implementation of the paradigm laid out in this chapter are given in Appendix~\ref{sec:software-implementation}.

\subsubsection{Handling Gate Cancellation}
\label{sec:cancel}

During the process of preparing a quantum circuit to run on a quantum device, one typically wishes to optimize the specific choice of gate operations, usually in the interest of decreasing the overall circuit length. While a full discussion of all possible optimizations is beyond the scope of this work, we mention one concrete case here as an example indicating how to apply our methods in the context of such optimizations.

\begin{figure}
   \centering
   \includegraphics[width=\columnwidth]{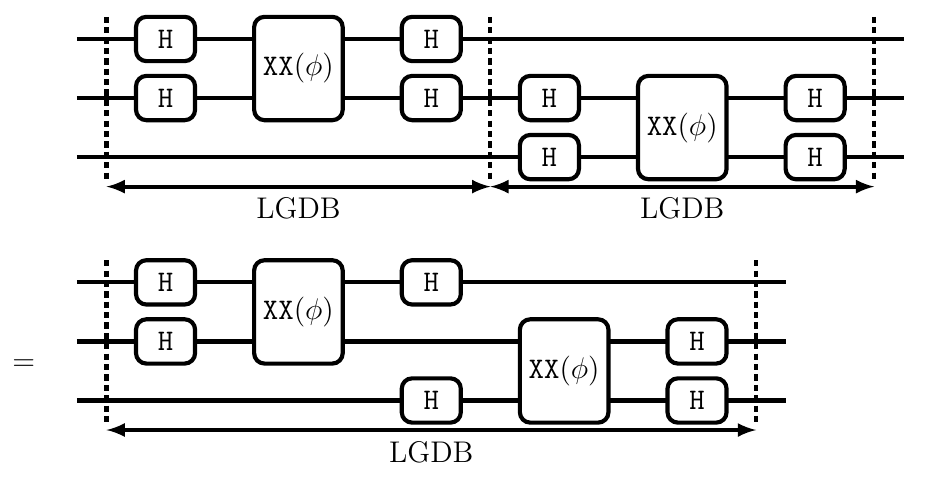}
   \caption{Two \gate{ZZ} Ising interactions between overlapping pairs of qubits, where the interactions are each decomposed into \gate{XX} interactions and surrounding Hadamard gates that transform between the $X$ and $Z$ basis. Since Hadamard gates are self-inverse, two of the Hadamard gates could be canceled. However, this would cause two LGDBs to be merged into one. For larger circuits one could in a similar fashion create potentially huge decomposition blocks.}
   \label{fig:cancel}
\end{figure}

In particular, we imagine a circuit like the one shown in Figure~\ref{fig:cancel}, which is designed to implement some \gate{ZZ} pair interactions between overlapping pairs of qubits.
We take the \gate{ZZ} gates to be realized by \gate{XX} interactions with surrounding Hadamard gates, in order to implement the appropriate basis change. In this example, two of the Hadamard gates would cancel each other. This is often the case in such decompositions, where the action of two or more gates counteract each other; hence, they need not be included in the final circuit which is implemented on the actual quantum device. One consequence of canceling these two gates, however, is that two separate LGDBs have now been effectively fused into one single LGDB, which involves all three qubits. Such a fusion of LGDBs has the potential to become computationally expensive, especially for larger circuits, for the reasons mentioned when discussing the case of \gate{SWAP} blocks.

\begin{figure}
   \centering
   \includegraphics[width=\columnwidth]{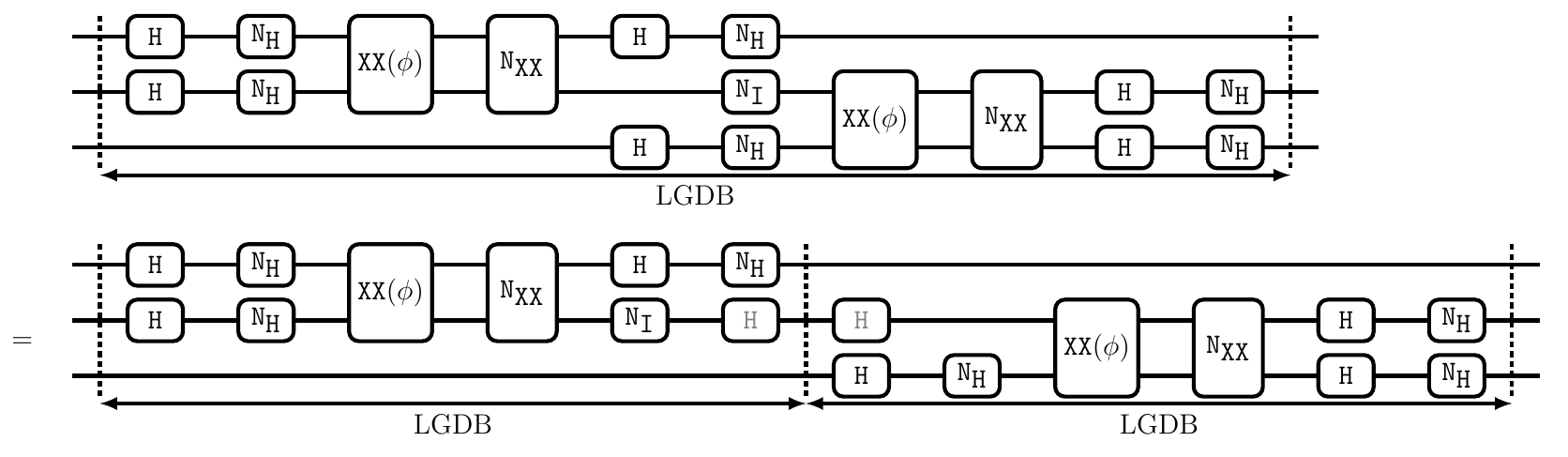}
   \caption{The circuit of Figure~\ref{fig:cancel} without the unneeded \gate{H} gates, but with noise terms for each gate, including a noise term for the middle qubit idling, while the outer qubits perform a Hadamard each. One can reintroduce two noise-free (fictitious) Hadamard gates to recover the original LGDBs.}
   \label{fig:cancel_noisy}
\end{figure}

Fortunately, the solution to this problem is fairly simple. After gate cancellations and noise terms have been accounted for, one must simply reintroduce the original gates into the circuit as fictitious gates, as shown in Figure~\ref{fig:cancel_noisy}. In the circuit of the previous Figure~\ref{fig:cancel} with the canceled Hadamard gates, we introduce noise terms after each gate. We also account for a potential noise term that accumulates while the middle qubit idles during the time in which the other two execute the \gate{H} gates.
This is a representation of what would happen on the actual hardware when the circuit is executed without the additional Hadamard gates (which is of course different from an execution including the additional gates).

This representation is equal to a circuit with two additional noise-free (fictitious) \gate{H} gates, as indicated in Figure~\ref{fig:cancel_noisy} in gray.
Reintroducing these imaginary gates allows, however, to recover the original LGDBs.
We can now proceed with the usual analysis of this circuit, without the need to consider a LGDB which spans the entire system.
It is important to emphasize, however, that these fictitious gates must be placed directly next to each other, without any noise terms in between, so that their product is indeed still equal to the identity gate.

We see from the example above that we can handle this case of gate optimization by effectively reverting the circuit back to its original form, through the use of fictitious gates which of course do not posses their own noise terms. For larger circuits with more involved gate optimizations, it is therefore useful to save the information regarding the original form of the circuit before gate optimization, since it may in general be difficult (or infeasible) to recover the original circuit from the optimized one. We emphasize, however, that it is not strictly necessary to revert the circuit back to exactly its original form - any form of the circuit which helps avoid computational difficulties due to LGDBs which span the whole system would be satisfactory, which may or may not be simpler than the original circuit before optimization.

\subsection{Scaling of Errors}
\label{sec:scale}

We have thus far ignored the scaling of errors in our analysis. Here we briefly comment on this, while leaving a more detailed discussion to Appendix \ref{sec:appError}.

There are two main sources of error to consider. Again, we will assume that there are some characteristic $\phi$ and $\mu$ which describe the coherent and incoherent terms throughout the circuit, respectively. The first source of error results from only summing to zero order in the BCH expansion. For the coherent dynamics, we find that this introduces an error of order $\phi$, while for the noisy portion of the dynamics, we find an error of order $\mu$. We note that each of these corrections is of order $\phi$ smaller than their dominant zero order contributions. 

The second source of error results from naively commuting noise terms past small angle gates in SWAP blocks. We find that the error in this case is a contribution to the noisy dynamics, which is again a factor of $\phi$ smaller than the dominant dynamics. 

We note that when higher-order Suzuki-Trotter decompositions are utilized, additional care is required when considering the dominant sources of error. For the case of a second-order Suzuki-Trotter decomposition, we find that the errors introduced through our approximate noise model will be small compared with the errors introduced through Trotterization so long as
\begin{equation}
\mu \ll \phi^{2}.
\end{equation}

\section{Numerical Verification}
\label{sec:numerical-analysis}

\begin{figure}
   \centering
   \includegraphics[width=\columnwidth]{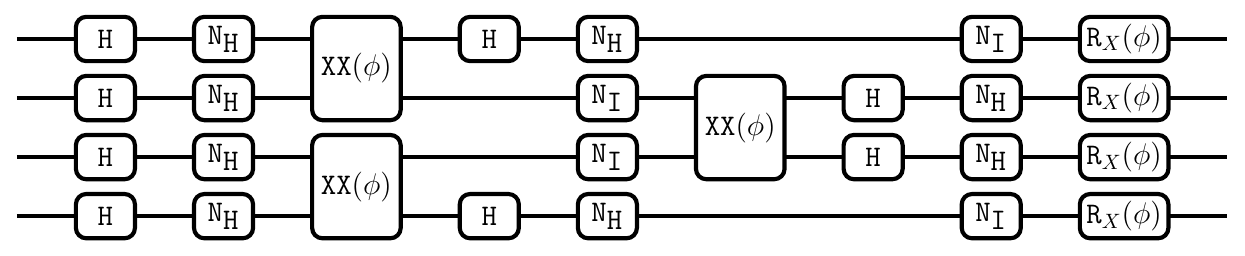}
   \caption{An example circuit that implements a Trotter step of the time evolution under the example Ising Hamiltonian~\ref{eq:ising-example-Hamiltonian}. We defined $\phi = 2J\tau = 2g\tau$, with the parameters $J$ and $g$ from the Hamiltonian and the Trotter step size $\tau$. We assumed noise events happening when applying the large angle rotations for the Hadamard gate on a qubit ($\gate{N}_\gate{H}$), and when a qubit is idling during the time Hadamard gates are applied elsewhere ($\gate{N}_\gate{I}$).}
   \label{fig:ex}
\end{figure}

In order to verify the accuracy of the formulas we have derived in the previous section, we have tested the time evolution of our noisy algorithm model against the exact time evolution of the quantum circuit from which it was derived, for a variety of example circuits. We have found good agreement in all of the cases we have tested. We display the results of one such test here. We will focus on the example circuit displayed in Figure~\ref{fig:ex}, intended as a toy model for a circuit which could be analyzed with our methods. Such a circuit may correspond to the digital simulation of the time evolution of the transverse field Ising model,
\begin{equation}\label{eq:ising-example-Hamiltonian}
H = J\sum_{\langle i j \rangle} \sigma^{z}_{i} \sigma^{z}_{j} + g \sum_{i} \sigma^{x}_{i},
\end{equation}
defined on a chain of four spins. We assume that small angle Ising terms and small angle single qubit rotations along the X-axis are natively available on the device hardware. However, in contrast, we assume that only special large angles (for example, multiples of $\pi / 2$) are available for rotations and interactions along the Y and Z directions. Therefore, the use of Hadamard gates is necessary in order to implement rotations and interactions along more than one axis, which could be placed around either the Ising terms, or the single qubit terms. In this example, they are placed around the Ising terms, in order to accomplish an Ising interaction along the Z axis. 

We further assume a toy model for the discrete noise term which follows a Hadamard gate, given in \cite{nap}. This discrete noise term is derived for a Hadamard gate which is composed of several elementary rotations, under the assumption that these rotations are being applied to a qubit affected by independent damping noise which is not altered by the introduction of coherent dynamics (we will in fact adopt this assumption of constant damping noise for our example circuit as a whole). If the time required to implement a gate is roughly proportional to its gate angle, we can assume that the noise accumulated during the implementation of the small angle gates (single qubit rotations and Ising interactions) is negligible compared with the noise following a Hadamard gate. However, we do account for the damping noise which accumulates on idling qubits while other qubits have Hadamard gates applied to them (we assume that gates which appear in the same vertical block are implemented in parallel on the hardware).

\begin{figure}
   \centering
   \includegraphics[width=\columnwidth]{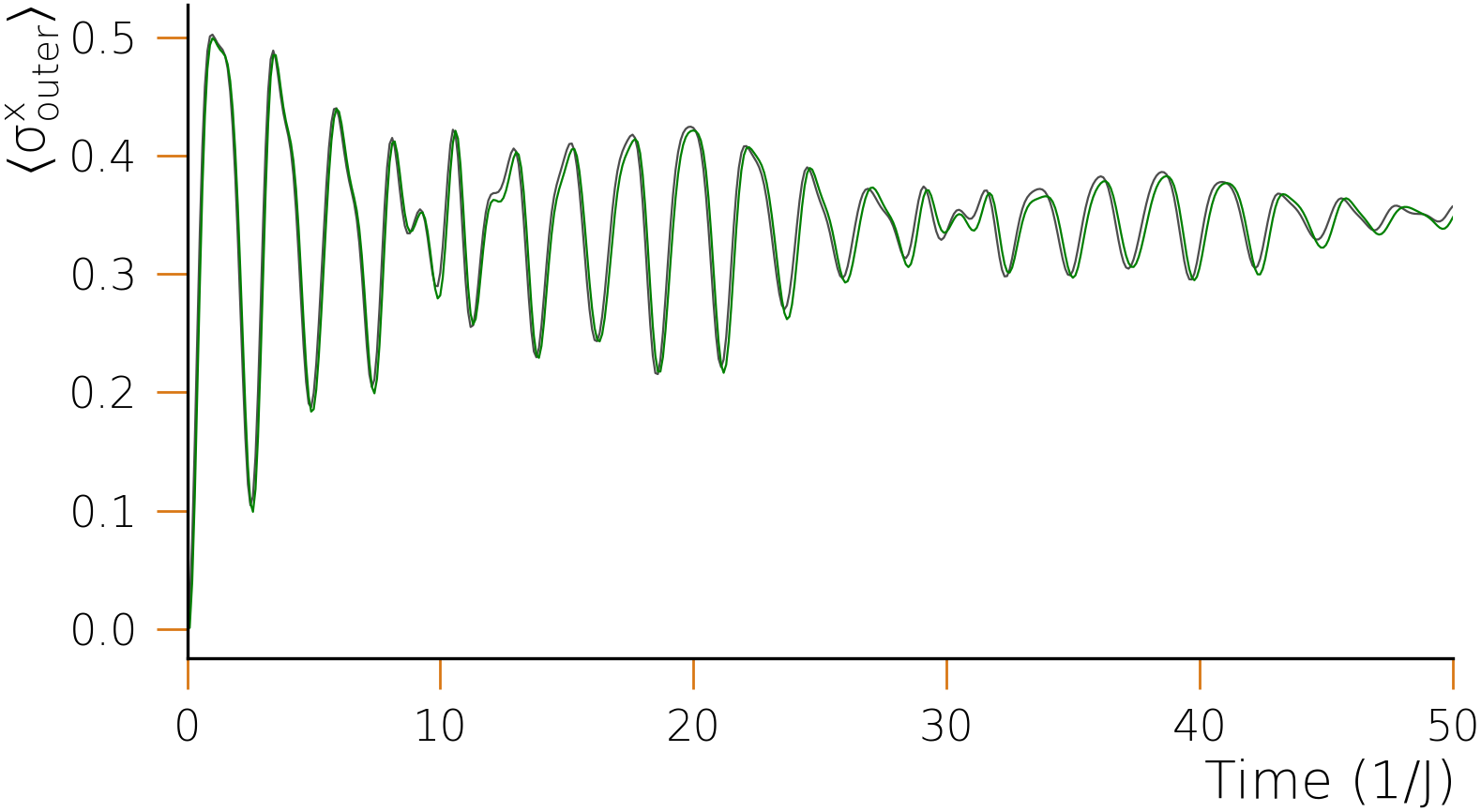}
   \caption{The comparison between the exact circuit evolution (green) and the time evolution under the noisy algorithm model (black).}
   \label{fig:plotA}
\end{figure}

In Figure~\ref{fig:plotA} we display the results of such a time evolution comparison. Here we take $J\tau = g\tau = 0.1$, with a (uniform) noise strength given according to $\gamma_{\text{damp}}t_{H} = 10^{-3}$, where $t_{H}$ is the time required to implement a Hadamard gate (for such a parameter choice, the single qubit rotations and Ising interactions possess gate times roughly 25 times smaller than the gate time of the Hadamard gate given in \cite{nap}, justifying our decision to neglect the noise following these small angle gates). Shown is the ``exact'' time evolution according to the operations and noise terms precisely as they appear in the circuit, as well as a time evolution according to the noisy algorithm model. The strong agreement validates the results of our analysis. Here we choose to display the time evolution of $\sigma^{x}$ on either of the two outer sites, starting from an initial state in which all spins are polarized along the positive Z axis, although all observables in the system display a similar level of agreement, for any initial state. We note that our choice of parameters for this example circuit does indeed obey the relationship
\begin{equation}
\mu \ll \phi^{2} \ll \phi.
\end{equation}

\begin{figure}
   \centering
   \includegraphics[width=\columnwidth]{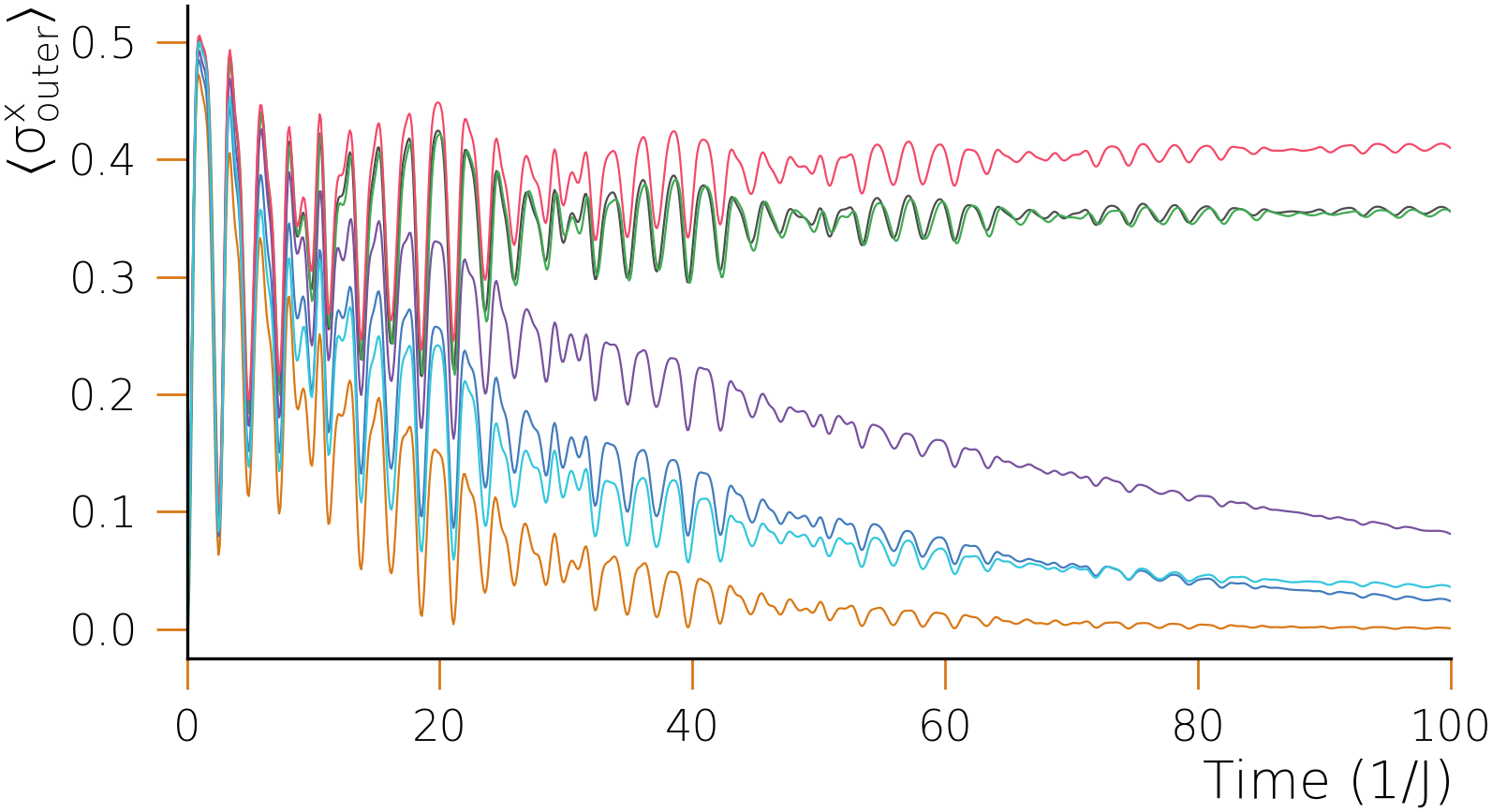}
   \caption{The time evolution, compared against several other potential noise models. The effective noise models which assume independent damping (light blue), dephasing (purple), and depolarizing (dark blue) all show poor agreement with the actual circuit dynamics. The same can be said for the global depolarizing channel (gold), and the effective model in which we attempt to compute the noisy algorithm model, but neglect to account for the effects of commuting noise terms past large angle gates (red). Only the noisy algorithm model (black) shows good agreement with the actual circuit dynamics (green).}
   \label{fig:plotB}
\end{figure}

In addition to displaying the time evolution under the noisy algorithm model, in Figure~\ref{fig:plotB} we see a comparison with several other models which one might assume for the effective noise occurring during the simulation. In particular, we compare our results against models in which we assume that the effective noise is equivalent to uniform, uncorrelated damping, dephasing, or depolarizing noise. We also display the time evolution under a model in which we assume that the effects of noise can be accounted for by simply applying a depolarization channel to the global density matrix after each Trotter step,
\begin{equation}
\rho \to (1-\lambda)\rho+\frac{\lambda}{\mathcal{D}}I,
\end{equation}
which is equivalent to time evolving the system with the Lindblad noise
\begin{equation}
\mathcal{L} \left [ \rho \right ] = \frac{\gamma_{DC}}{\mathcal{D}^{2}} \sum_{\alpha} \left [ A_{\alpha} \rho A_{\alpha} - \rho \right ]
\end{equation}
where the sum is over any complete orthonormal set of traceless operators $\left\{A_{\alpha}\right\}$, and
\begin{equation}
\lambda = 1 - e^{-\gamma_{DC} \tau}.
\end{equation}
Since this global depolarizing channel commutes with any coherent dynamics \cite{nap}, a discrete depolarizing channel being applied at the end of a Trotter step is equivalent to the Lindblad term above being applied concurrently with the Hamiltonian dynamics. Lastly, we show a comparison to the noisy algorithm model one would find when failing to properly account for the effects of commuting noise terms past large gates. In order to determine the appropriate noise strengths for the various alternate choices of effective noise model, we demand that the trace of the rate matrix (which is equal to the sum of its eigenvalues, and independent of basis \cite{nap}) should remain the same as the noisy algorithm model.

In all such cases, we see very poor agreement with the actual circuit dynamics. Not only are the amplitudes of the curves noticeably different, but the steady states are clearly not in agreement. This motivates the need for the careful noise analysis we have performed in the previous section. Note that for several of the alternative noise models, it would appear as though the qualitative features of the plots are similar, with only deviations in the amplitude being present, and so perhaps the curves could be brought into agreement by adjusting the overall noise strength of the models (rather than simply matching the trace of the rate matrix to that of the noisy algorithm model). However, this in fact does not work - manually adjusting the noise rates in such a way will also lead to a qualitative change in the dynamics, such that when the two curves are made to overlap, the qualitative features no longer agree. Additionally, even if we were to attempt to adjust the overall noise strength in this way, this is not able to account for the difference in the steady states.

\section{Conclusion}
\label{sec:conclusion}

In this work we have presented a method for analyzing the effects of noise in quantum circuits, specifically those which are designed to simulate the time evolution of quantum spin systems through Trotterization. Our method describes the effects of noise through the use of an effective noise model, the \textit{noisy algorithm model}, the time evolution of which provides a good approximation to the actual dynamics of the noisy quantum circuit. While the results presented here have considered the case of time-independent Hamiltonians, the generalization to the time-dependent case is straight-forward.

An obvious generalization of our work would be the extension to Hamiltonians with degrees of freedom other than spins - for example, fermionic degrees of freedom, implemented on a quantum device through the Jordan-Wigner transformation. The philosophy of our analysis, according to which noise terms are shuffled around in a circuit until only small-angle gates and noise terms remain, should still be valid in such a case. The primary consideration, however, would be whether the resulting noise model would have any physical interpretation, after the spin degrees of freedom are transformed back to fermionic degrees of freedom~\cite{reiner_effects_2018}. Similarly, it would be interesting to consider how our work could be extended beyond the scope of standard Trotterization, for example for investigating optimized circuits for quantum simulation~\cite{PhysRevResearch.5.023146, nam_automated_2018}. We leave further investigation of these matters as an open question.

The noisy algorithm model may not only help to understand the extent to which the result of a digital quantum simulation may differ from the intended calculation: Potentially, one could use this framework to find open quantum systems of interest that may be faithfully simulated on a noisy quantum device. That would be the case, if one tailors the circuit, as well as the noise of the device~\cite{barreiro_open-system_2011,rost_simulation_2020}, such that the noisy algorithm model coincides with that open quantum system~\cite{leppaekangas2023quantum}. Simulating open quantum systems is an active field of research~\cite{hu_quantum_2020, schlimgen_quantum_2021, jo_simulating_2022}, hence, offering a new computational tool could be highly beneficial~\cite{verstraete_quantum_2009,chen_local_2025,zhan_rapid_2025}. However, studying this question is out of the scope of this manuscript.

We hope the results presented in this work will allow for further progress in the field of digital quantum simulation.

\acknowledgments{
This work received funding from the European Union’s Horizon program with numbers 899561 (AVaQus) and 101046968 (BRISQ). We thank Eric Dzienkowski, Sebastian Fischetti, and Brayden Ware for helpful discussions.
}

\newpage

\bibliographystyle{unsrtnat}
\bibliography{Sections/refs}

\newpage
%~\newpage

\appendix

\section{Software Implementation Details}
\label{sec:software-implementation}

In this appendix, we explain in more detail how we implemented the software to calculate the effective Lindbladian of a noisy quantum algorithm. The tool is built upon two open-source software packages from HQS Quantum Simulations:
Firstly, \textit{struqture}~\cite{struqture}, which is used to represent Hamiltonians and Lindblad operators in our code.
Furthermore, \textit{qoqo}~\cite{noauthor_github_nodate}, the package for representing quantum circuits and programs, with backend options for running them on quantum devices, but also a simulator backend for conventional hardware relying on QuEST~\cite{noauthor_quest_nodate}.

In qoqo, we have the particular freedom to define on the circuit level any kind of Lindblad noise acting during the application of the quantum program. Hence, we can produce generic noisy quantum algorithms that conform with our noise assumption of Section~\ref{sec:noise-assumptions}. Furthermore, the notion of LGDBs from the main text is also included in qoqo, where they are called \textit{decomposition blocks}. Hence, we have all the ingredients to build a tool to calculate the noisy algorithm model:

As an input, the code takes the noisy quantum circuit (i.e., including noise gates) implementing a Trotter step of the time evolution of a Spin system for which we want to find the noisy algorithm model would effectively be simulated on noisy hardware. Also as an input, the tool will receive the Trotter step size which is important for the correct rescaling of the Lindblad terms at the end.

In the noisy quantum circuit, every portion needs to be wrapped in decomposition blocks to signal that the respective gate sequences within the blocks correspond to the application of a exponential that is parametrically small (i.e., comparable or smaller than the Trotter step size).
This is important information for the code to be able to commute noise through the circuit while staying within the error bars of the Trotter approximation.
Furthermore, the decomposition blocks contain all the information about swapping operations, and whether the qubit indices the noise acts on need to be remapped when commuting noise past the block, to deal with the issue highlighted in~\ref{sec:SWAP-handling}.

We now go through the circuit starting with the first decomposition block. For this block, we get the list of the qubits that are involved in this block, as well as the list of operations, in order of execution.
The code then goes through the list of operations, and finds the noise operations. For each noise operation, we build the Lindblad operator in a matrix form, and we convert the following gate operation to a physical operator. We can then commute the noise term past the gate operation, and we get a transformed noise term, where the unitary transformation is defined by the unitary matrix that represents the gate, as shown in Eq.~\ref{eq:noise-basis-transformation}. We perform this every time we get to a gate operation, and need to commute the noise term past it, until we reach the end of the decomposition block.

Afterwards, the transformed noise terms will be commuted past the following decomposition blocks, where the qubit indices of the noise terms are re\-mapped according to the remapping information of the blocks (stating if there are swapping operations contained in the block).

Finally, the noise terms are rescaled by the Trotter step size, as in the sum in Eq.~\ref{eq:noise-rescaling} or Eq.~\ref{eq:noise-resclaing-P}.

This procedure is repeated for the second, third, etc., decomposition block in the noisy quantum circuit until we have treated every decomposition block in the circuit. Finally, the noise terms are added up following Eq.~\ref{eq:noise-rescaling} and Eq.~\ref{eq:noise-resclaing-P}.

The output of our tool is then the sum of all the commuted though, and rescaled, noise terms. This corresponds to the Lindblad noise operator of the noisy algorithm model.

A software implementation of this procedure can be found within the HQS cloud offering~\cite{HQStage}.
Specifically, the core functionalities exist in the \textit{HQS Quantum Libraries} module, which are furthermore utilized in the \textit{HQS Noise App}.

\section{Error Analysis}
\label{sec:appError}

When deriving the effective model being simulated by a given circuit, it is important to also consider the dominant sources of error in the approximations being made. Before doing so, however, we must first clarify some notation. We assume that the Hamiltonian we wish to simulate possesses terms with some characteristic (coupling) scale $J$. We define the product of this coupling with the Trotter step size as
\begin{equation}
\phi = 2 J \tau.
\end{equation}
For a super-operator corresponding to a coherent gate term, we define the scaled super-operator as
\begin{equation}
\mathcal{L}_{G} = 2 J \overline{\mathcal{L}}_{G} 
\end{equation}
Thus, the quantity appearing in the exponential corresponding to a coherent gate is
\begin{equation}
\tau \mathcal{L}_{G} = \phi \overline{\mathcal{L}}_{G} 
\end{equation}
Likewise, we assume a characteristic noise scale $\gamma$ and characteristic gate time $t_{G}$, and define
\begin{equation}
\mu = \gamma t_{G}.
\end{equation}
The scaled noise operator for a given gate is defined according to
\begin{equation}
t_{G} \mathcal{L}^{G}_{N} = \mu \overline{\mathcal{L}}^{G}_{N}
\end{equation}
We also define a scaled version of the effective super-operator according to,
\begin{equation}
\tau \mathcal{L}_{\text{eff}} = \phi \overline{\mathcal{L}}_{\text{eff}}
\end{equation}
In our analysis, we will be interested in understanding corrections to $\overline{\mathcal{L}}_{\text{eff}}$, thus allowing us to more easily analyze the scaling of various errors. We note that with these definitions, our zero order result can be written
\begin{equation}
\overline{\mathcal{L}}_{\text{eff}} \approx \sum_{X} \overline{\mathcal{L}}_{X} + \frac{\mu}{\phi}\sum_{g}\overline{\mathcal{L}}^{g}_{N}.
\end{equation}

\subsection{Higher Order Corrections in the BCH Formula}

Thus far, when adding together all of the small parameter terms in a circuit, we have only carried out the BCH expansion to zero order. Now we would like to discuss the error which has been introduced by this approximation. When converting a product of exponentials into a single exponential,
\begin{equation}
\prod_{i} e^{A_{i}} = e^Z,
\end{equation}
the BCH formula gives, to first order,
\begin{equation}
Z = \sum_{i} A_{i} + \frac{1}{2}\sum_{i < j} \left [ A_{i}, A_{j} \right ],
\end{equation}
where $i < j$ when $A_{i}$ appears to the left of $A_{j}$ in the product. The first term is of course the zero order contribution we have already considered, while the second term represents a first order correction. Since the circuits we consider will consist of small angle gates and noise terms, there are three types of commutators which will appear: gate terms with gate terms, gate terms with noise terms, and noise terms with noise terms. We consider each of these cases in turn.

First, it is clear that the commutator of two gate terms will produce another gate term. In particular,
\begin{equation}
\left [ \overline{\mathcal{L}}_{X}, \overline{\mathcal{L}}_{Y} \right ] = \overline{\mathcal{L}}_{W},
\end{equation}
where
\begin{equation}
\overline{\mathcal{L}}_{Q} \equiv -i \left [\overline{h}_{Q}, \cdot \right ]
\end{equation}
and
\begin{equation}
\overline{h}_{W} \equiv -i \left [ \overline{h}_{X}, \overline{h}_{Y} \right ].
\end{equation}
This is of course the usual correction to the coherent dynamics which results from the Trotterization of the circuit, and would be present even in the absence of noise. Since each coherent term comes with a factor of $\phi$ in the exponential, this leads to a correction to $\overline{\mathcal{L}}_{\text{eff}}$ which scales as
\begin{equation}
\Delta \left ( \phi \overline{\mathcal{L}}_{\text{eff}} \right ) \sim \phi \cdot \phi ~ \Rightarrow ~ \Delta \overline{\mathcal{L}}_{\text{eff}} \sim \phi
\end{equation}

Next, we consider commutators between gate terms and noise terms, which take the form of another noise term \cite{nap}. Since each gate term carries a factor of $\phi$, and each noise term carries a factor of $\mu$, such a term leads to a correction of order
\begin{equation}
\Delta \left ( \phi \overline{\mathcal{L}}_{\text{eff}} \right ) \sim \phi \cdot \mu ~ \Rightarrow ~ \Delta \overline{\mathcal{L}}_{\text{eff}} \sim \mu
\end{equation}
Since we assume $\mu \ll \phi$, in absolute terms, this represents a correction to the dynamics which is small compared with the previous correction we have found. If we instead focus only on the noisy part of the dynamics, such a term represents a correction to the noisy dynamics which is a factor $\phi$ smaller than the order zero noisy dynamics. This relative scale is the same as for the correction to the coherent dynamics.

Lastly, we consider noise terms with noise terms. Such commutators have a mixed form - they do not necessarily take the form of a coherent correction or a noisy correction. Since each of these terms carries a factor of $\mu$, we find a correction which scales as
\begin{equation}
\Delta \left ( \phi \overline{\mathcal{L}}_{\text{eff}} \right ) \sim \mu \cdot \mu ~ \Rightarrow ~ \Delta \overline{\mathcal{L}}_{\text{eff}} \sim \mu^{2} / \phi
\end{equation}
Again, since we assume $\mu \ll \phi$, we have in this case
\begin{equation}
\Delta \overline{\mathcal{L}}_{\text{eff}} \sim \mu^{2} / \phi \ll \left ( \mu \phi \right ) / \phi = \mu  \ll \phi,
\end{equation}
and thus we find a correction which is significantly smaller than either of the two previous corrections.

Thus, we find the two most important corrections to the dynamics to be a coherent correction of order $\phi$, and a noisy correction of order $\mu$, each of which is a factor of $\phi$ smaller than their respective zero order contributions.

\subsection{Corrections appearing in SWAP Blocks}

When handling SWAP blocks, we have not thus far accounted for the effects of commuting a noise term past a small angle gate. Here, we wish to analyze the error introduced through this approximation.

We know that the modification of a noise term resulting from commuting it past a gate involves a correction to its underlying rate matrix which is given according to
\begin{equation}
\Gamma^{N} ~ \to ~ \Gamma^{P} ~ \equiv ~ M\Gamma^{N}M^\dagger
\end{equation}
where
\begin{equation}
M_{mn} = \frac{1}{\mathcal{D}}\text{Tr} \left [ A^{\dagger}_{m}U A_{n} U^{\dagger} \right ].
\end{equation}
When the gate in question is a small angle gate, we can approximate
\begin{equation}
U \left ( 2 J \tau \right ) = \exp \left ( - 2 i J \overline{h} \tau \right ) \approx I - i \phi \overline{h} + \mathcal{O} \left ( \phi^{2} \right )
\end{equation}
Our transformation therefore becomes
\begin{equation}
M_{mn} = \delta_{mn} + i \phi\Omega_{mn},
\end{equation}
where
\begin{equation}
\Omega_{mn} \equiv \frac{1}{\mathcal{D}}\text{Tr} \left [ A^{\dagger}_{m} A_{n} \overline{h} \right ] - \frac{1}{\mathcal{D}}\text{Tr} \left [ A^{\dagger}_{m} \overline{h} A_{n} \right ]
\end{equation}
Thus, we find in this case
\begin{equation}
\begin{split}
\Gamma^{N} ~ \to ~ \Gamma^{P} ~ & \approx ~ \left (I + i \phi \Omega \right ) \Gamma^{N} \left (I - i \phi \Omega \right ) \\
& \approx ~ \Gamma^{N} + i \phi \left [ \Omega, \Gamma^{N} \right ]
\end{split}
\end{equation}

We have thus introduced an error of order $\phi$ when neglecting the effects of commuting a noise term past a small angle gate, which is the same order as neglecting higher order terms in the BCH expansion. As a concrete example, if we consider a single qubit, with the basis of traceless operators equal to the usual Pauli operators, and the scaled Hamiltonian equal to
\begin{equation}
\overline{h} = -\frac{1}{2}\sigma^{x}
\end{equation}
then we have
\begin{equation}
\Omega = \lambda_{7},
\end{equation}
where $\lambda_{7}$ is the seventh Gell-Mann matrix.

\subsection{Higher-Order Trotter Decompositions}

\begin{figure}
   \centering
   \includegraphics[width=\columnwidth]{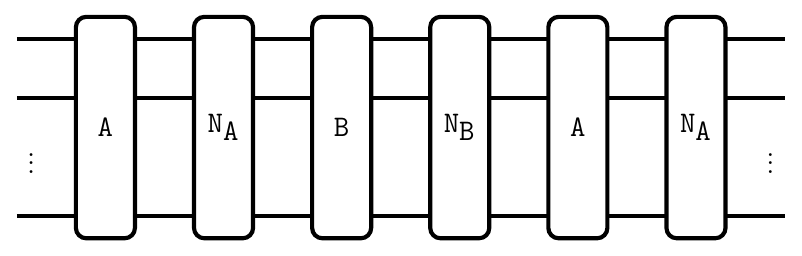}
   \caption{A very rough schematic of a second-order decomposition, with noise.}
   \label{fig:SOST}
\end{figure}

In many situations, it is desirable to reduce the errors introduced through Trotterization by introducing a second order Trotter decomposition,
\begin{equation}
U \left ( \tau \right ) \approx \prod^{Q}_{X=1}  U_{X} \left ( \tau / 2\right ) \prod^{1}_{X=Q}  U_{X} \left ( \tau / 2\right )
\end{equation}
where $Q$ is the number of Hamiltonian terms. In other words, a trotter step of time $\tau$ is implemented through a forwards sequence of gates, followed by the same sequence of gates in reverse. An example of such a circuit is shown in Figure~\ref{fig:SOST}. In such a case, each product $U_{X}$ may need to be decomposed into a sequence of gates, each of which possessing its own noise term. Again, with a suitable set of manipulations, this circuit can be brought into the form of a sequence of SAGs, with noise terms interspersed among them. Our previously found formula for the effective model being simulated is still valid in this case. However, we must now think more carefully about the dominant sources of error.

By design, a second order Trotter decomposition is constructed such that any correction to the Hamiltonian will scale as $\mathcal{O} \left ( \phi^{2} \right )$. The symmetric arrangement of terms implies that every commutator appearing in the expression for the first order correction is paired with another commutator in reverse order. Examining Figure~\ref{fig:SOST}, it is also clear that the same should hold true for the noise terms. If we assume that a given gate is always followed by the same noise term, and if we assume that a given term $U_{X}$ is always decomposed using the same set of available gates, then after manipulating our circuit into one which possesses only SAGs and noise terms, the noise terms themselves should also be arranged symmetrically, and thus any first order contribution from commutators of noise terms with noise terms should also vanish (if these assumptions are not valid, then any asymmetry in the arrangement of noise terms will result in first order corrections which again scale as $\mathcal{O} \left ( \mu^{2} / \phi \right )$).

\begin{figure}
   \centering
   \includegraphics[width=\columnwidth]{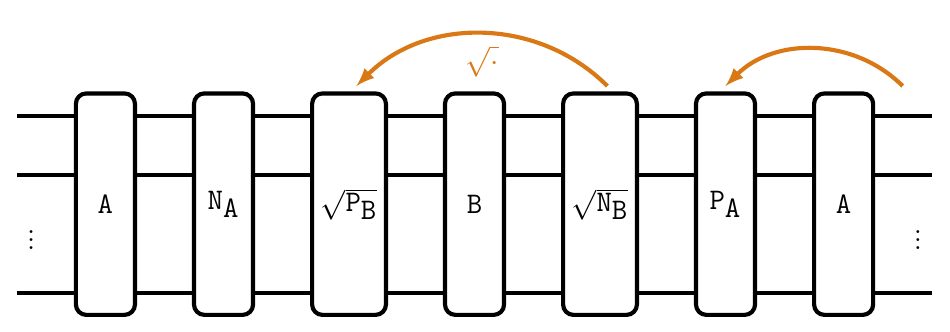}
   \caption{The idea of trying to symmetrize noise. The noise after gate B is split in half, with one half being commuted over. The noise after the second A gate is also commuted over. While this appears superficially symmetric, and as if the noise terms should now cancel, there are in fact order $\theta$ corrections to the noise when commuting them, and so the cancellation is not perfect. This results in the same amount of error as if we had simply added them up in the original arrangement.}
   \label{fig:splitNoise}
\end{figure}

However, no such symmetry exists between gate terms and noise terms. In general, since we model noise terms as following gate terms, there will be imperfect cancellation among the commutators of gate terms with noise terms, leading again to a correction of order $\mu$. One might imagine that this could be remedied by modeling the noise differently in the second half of the circuit. If one models the noise terms as appearing before their corresponding gates in the second half of the circuit, the arrangement of gate terms and noise terms would again appear to be symmetric. However, when modeling the effects of noise, the actual evolution of the qubit register must remain the same (since we are only describing the effects of noise, and not changing anything about the actual circuit or hardware), and so if we alter the nature of our noise model to place the noise terms before, rather than after, a given gate, then the nature of the noise term itself must also change, in order to compensate. In particular, we know that commuting a noise term past a SAG results in corrections of order $\phi$ which, when combined with the overall effective noise scale $\mu / \phi$, will again lead to corrections of order $\mu$. Thus, even though noise and gate commutators appear in symmetric pairs, there is still an asymmetry  of order $\mu$. This idea is demonstrated in Figure~\ref{fig:splitNoise}.

In such a second order circuit, we thus now have corrections to the noisy dynamics which scale as $\mu$, and corrections to the coherent dynamics which scale as $\phi^{2}$. We note that so long as $\mu \ll \phi^{2}$, the error in the noisy dynamics will still be small compared with the error introduced through Trotterization.

\end{document}